\documentstyle[prl,aps,epsf,floats]{revtex}
\begin{document}
\draft
\twocolumn[\hsize\textwidth\columnwidth\hsize\csname @twocolumnfalse\endcsname
\title {Band Structure of the Jahn-Teller Polaron from
        Quantum Monte Carlo} 

\author{P.\,E.\,Kornilovitch}
\address{
Blackett Laboratory, Imperial College, Prince Consort Road, London SW7 2BZ, UK \\
$^*$Hewlett-Packard Laboratories, 3500 Deer Creek Road, Palo Alto, 
California, 94304
}

\date{\today}
\maketitle
\begin{abstract}

A path-integral representation is constructed for the Jahn-Teller
polaron (JTP).  It leads to a perturbation series that can be summed
exactly by the diagrammatic Quantum Monte Carlo technique.
The ground-state energy, effective mass, spectrum and density
of states of the three-dimensional JTP are calculated with no
systematic errors.  The band structure of JTP interacting with 
dispersionless phonons, is found to be similar to that of the 
Holstein polaron. The mass of JTP increases exponentially 
with the coupling constant.  At small phonon frequencies, the spectrum 
of JTP is flat at large momenta, which leads to a strongly distorted 
density of states with a massive peak at the top of the band.     

\end{abstract}
\vskip2pc]
\narrowtext

The renewed interest in the Jahn-Teller polaron (JTP) problem was 
sparked by its possible relevance to the colossal magnetoresistance 
phenomenon in manganese oxides \cite{Millis,Mueller,Alex&Brat,Allen,Vasiliu}. 
Although cooperative properties of JTPs are of most importance,
the solution of the single-JTP problem is a necessary step towards
the comprehensive understanding of the physics of these and other 
materials with Jahn-Teller ions.
JTP also bears a significant fundamental interest and it has been a
subject of intense theoretical research throughout decades, see, 
e.g., \cite{Kanamori,Abragam,Kugel&Khomskii}. However, the full 
quantum-mechanical treatment of this problem is difficult, and 
no exact solution has been found. Even the single-site 
JTP does not allow analytical solution. The latter fact has so 
far prevented the development of a strong-coupling perturbation 
theory, so successful in the theory of the Holstein polaron 
\cite{Lang&Firsov}. Numerical methods, such as exact diagonalization and
density-matrix renormalization group, face the potential difficulty
in dealing with a very large Hilbert space of the two phonon modes
which are the essential feature of the JTP. Quantum Monte Carlo studies
reported so far have treated the phonons only classically 
\cite{Dagotto}.

In this paper, a method is presented that allows the full quantum
mechanical solution of the JTP problem. The method combines algorithms
developed in the Monte Carlo studies of the Holstein 
\cite{DeRaedt,Kornilovitch} and Fr\"ohlich \cite{Prokofiev} polarons.
It is based on a path-integral representation of JTP partition function,
analytical integration over the phonon coordinates \cite{Feynman,DeRaedt}, 
and numerical integration over the electron coordinates. The latter is done 
exactly, with the Diagrammatic Quantum Monte Carlo technique \cite{Prokofiev}. 
The use of free boundary conditions in imaginary time \cite{Kornilovitch} 
allows calculation of the polaron effective mass and the entire spectrum.  
The method is formulated in continuous time and no systematic errors 
are introduced. As a result, the ground-state energy, mass, spectrum, 
and density of states of JTP are calculated exactly on an infinite 
lattice in the most difficult three-dimensional case. We found that
the properties of JTP are very similar to that of the Holstein polaron (HP). 
JTP is as heavy as HP. It also displays some {\em anomalous} properties,
again similar to HP. In particular, in the case of  dispersionless phonons 
and low phonon frequency, JTP has a flat spectrum at large momenta and a 
very distorted density of states. 

The simplest JTP is an electron that hops between doubly degenerate 
$e_g$ levels and interacts locally with a doubly degenerate phonon 
mode. The model Hamiltonian reads \cite{Kanamori}
\begin{eqnarray}
 H        & = & H_e + H_{ph} + H_{e-ph} ,
\label{one}   \\
 H_e      & = & -t \sum_{\langle {\bf nn'} \rangle} 
(  c^{\dagger}_{{\bf n} 1} c_{{\bf n}' 1}
 + c^{\dagger}_{{\bf n} 2} c_{{\bf n}' 2} ) ,
\label{two}   \\
 H_{ph}   & = & \sum_{\bf n} \left[ \frac{1}{2M} ( \hat p^2_{x_{\bf n}} + 
\hat p^2_{y_{\bf n}} ) + \frac{M\omega^2}{2} (x^2_{\bf n} + y^2_{\bf n}) \right],
\label{three} \\ 
H_{e-ph}  & = & - \kappa \sum_{\bf n} [ 
  (c^{\dagger}_{{\bf n}1} c_{{\bf n}2} +  
   c^{\dagger}_{{\bf n}2} c_{{\bf n}1}) x_{\bf n}  \nonumber \\ 
&  & + (c^{\dagger}_{{\bf n}1} c_{{\bf n}1} -  
        c^{\dagger}_{{\bf n}2} c_{{\bf n}2}) y_{\bf n} ]. 
\label{four}
\end{eqnarray}
Here $c^{\dagger}_{{\bf n}a}$ creates an electron on site {\bf n} on orbital 
$a$ ($a=1$ or 2 is the orbital index), $t$ is the intersite hopping matrix
element, $\langle {\bf nn'} \rangle$ denotes pairs of nearest neighbors; 
$x_{\bf n}$ and $y_{\bf n}$ are the two phonon displacements on the
{\bf n}th site, $\hat p_{x_{\bf n},y_{\bf n}} = 
-i\hbar \partial_{x_{\bf n},y_{\bf n}}$; $M$ is the ionic mass; and $\kappa$ 
is the interaction parameter with dimensionality of force. The energy of
the atomic level is set to zero. The model is 
parameterized by the phonon frequency $\omega$ and by the dimensionless 
coupling constant $\lambda = \kappa^2/(2M\omega^2 z t)$ where $z$ is the 
number of nearest neighbors. For the simple cubic lattice, considered
in this work, $z=6$. Two simplifications are introduced in the model
(\ref{one})-(\ref{four}). First, the electron hopping is isotropic, and  
is diagonal in the orbital index. The latter leads to any change of orbital index 
taking place only when an $x$-phonon is emitted or absorbed. Averaging over
the phonons eliminates all the amplitudes with any odd total number of
emission and absorption acts. As a result, the off-diagonal in orbital
index elements of the polaron density matrix are identically zero. One
can show that such matrix elements determine the splitting between
the two polaron bands. The conclusion is that in the model
(\ref{one})-(\ref{four}), the two polaron bands are degenerate in
the whole Brillouin zone, and there exists only one ground-state dispersion 
relation $E_{\bf P}$. The second simplification is the absence of the phonon
dispersion. It is assumed that each site is surrounded by its own 
vibrating cluster and the neighboring clusters share no common ions.
Such a choice was made to bring the model close to the Holstein model
\cite{Holstein}. The difference between the two models lies exclusively
in the interaction term (\ref{four}). This fact will enable us to make
a reasonable comparison of JTP and HP. One should add that the full
Hamiltonian (\ref{one}) is invariant under the continuous transformation:
rotation in the plane $(x,y)$ by an angle $\phi$ (the same for
all sites) and simultaneous rotation in the plane $(c_1,c_2)$ by 
$\phi/2$. The invariance implies the equivalence of the $x$ and
$y$ modes in the sense that the elastic and interaction energies
associated with the two modes must be equal. The number of excited phonons
are equal too. These general results serve as independent checks of
numerical data.
 
Central in the proposed method is the path-integral representation for 
the density matrix $\rho$ of the model (\ref{one})-(\ref{four}). For a 
small imaginary-time interval $\Delta \tau = \beta/L$, where 
$\beta = (k_B T)^{-1}$ is the inverse temperature and $L \gg 1$,
one obtains, up to the first order in $\Delta \tau$:
\begin{eqnarray}
\rho(\Delta \tau) & = & \langle {\bf r}', a'; \{ x'_{\bf n} \}, \{ y'_{\bf n} \}|
e^{-\Delta \tau H} | {\bf r}, a; \{ x_{\bf n} \}, \{ y_{\bf n} \} \rangle  
                                                      \nonumber \\
                  & = & [ \delta_{\bf r r'} \delta_{a a'}
 + \Delta \tau \kappa \delta_{\bf r r'} \delta_{a \bar a'} x_{\bf r} \nonumber \\ 
&  & + \Delta \tau t \delta_{a a'} \sum_{\langle {\bf nn'} \rangle} 
\delta_{{\bf r},{\bf r'}+{\bf n}-{\bf n}'} ] e^{A^{\Delta \tau}_{ph}} ,
\label{five}
\end{eqnarray}
\begin{eqnarray}
A^{\Delta \tau}_{ph} = &   & \kappa \Delta \tau y_{\bf r} 
      ( \delta_{a 1} - \delta_{a 2} )           \nonumber \\ 
                       & - & \sum_{\bf n} \left\{  \frac{M}{2\hbar^2(\Delta \tau)} 
[ (x_{\bf n}-x'_{\bf n})^2 + (y_{\bf n}-y'_{\bf n})^2 ] \right. \nonumber \\
   &  & + \left. (\Delta \tau) \frac{M\omega^2}{2} ( x^2_{\bf n} + y^2_{\bf n} ) 
 \right\}  ,
\label{six}
\end{eqnarray}
$A^{\Delta \tau}_{ph}$ being the phonon action. Here $\bar a = 1, (2)$ 
when $a = 2, (1)$, i.e. $\bar a$ is `not' $a$. The full density matrix is obtained by 
multiplying $\rho(\Delta \tau)$ by itself $L$ times, introducing integration over 
the internal coordinates, and taking the $L \rightarrow \infty$ limit. 
In doing so, we do {\em not} impose the periodic boundary conditions in imaginary
time but rather leave initial ${\bf r}(0)$, $a(0)$ and final ${\bf r}(\beta)$, 
$a(\beta)$ coordinates of the electron arbitrary. [Except that
$a(\beta)=a(0)$ always holds.] The boundary conditions on the phonon coordinates 
are twisted, $x_{\bf n}(\beta) = x_{{\bf n}-{\bf r}(\beta)+{\bf r}(0)}(0)$, and 
the same for $y$, as described in \cite{Kornilovitch}. The next step is to integrate 
over $x_{\bf n}(\tau)$ and $y_{\bf n}(\tau)$. This is a Gaussian integration and it 
can be done analytically. Here one faces an important difference between the two 
phonon modes. Displacements $y_{\bf n}(\tau)$ interact with the electron density 
and do not change the electron coordinates. Therefore, for the oscillators 
$y_{\bf n}$, the electron is simply a source of external force, the value and 
sign of which depend on the position and orbital index of the electron, cf. the 
second term in Eq.~(\ref{four}). The problem reduces to uncoupled oscillators 
in a time-dependent external field. The integration over $y_{\bf n}(\tau)$ can be 
done by Feynman's methods \cite{Feynman} to yield the factor $e^{A_y}$ in 
the density matrix, where
\begin{eqnarray}
& A_y& [{\bf r}(\tau), a(\tau)]  \nonumber \\
& =  & \kappa^2 \int^{\beta}_0 \!\!\! \int^{\beta}_0 d\tau d\tau' 
G(\tau-\tau') [ \delta_{a(\tau), a(\tau')} - \delta_{a(\tau), \bar a(\tau')} ],
\label{seven}
\end{eqnarray}
\begin{eqnarray}
G(\tau-\tau') = \frac{\hbar}{2M\omega}  \left[ e^{-\hbar\omega|\tau-\tau'|} \right.  
\delta_{{\bf r}(\tau), {\bf r}(\tau')} \makebox[2.cm]{}  \nonumber \\
+ e^{-\hbar\omega(\beta-|\tau-\tau'|)} 
\left. \delta_{{\bf r}(\tau), 
{\bf r}(\tau')+{\rm sign}(\tau-\tau')\Delta {\bf r} } \right] ,
\label{eight}
\end{eqnarray}
is the retarded action induced by the $y$ mode. The second term in
the function $G(\tau-\tau')$ involves the net shift of the electron path 
$\Delta {\bf r} = {\bf r}(\beta) - {\bf r}(0)$ and the sign of the time
difference ${\rm sign}(\tau-\tau')=\pm 1$, \cite{Kornilovitch}. The simple 
form (\ref{eight}) of $G$ is valid
in the limit $e^{\beta \hbar \omega} \gg 1$ that is assumed hereafter. It is
important that $A_y$ is an explicit functional of the electron path and can be
calculated straightforwardly once the functions ${\bf r}(\tau)$ and $a(\tau)$ are
specified. Note, that the $y$-mode `favors' the same orbital index throughout the
whole electron path and `dislikes' orbital changes.

Unlike the $y$-mode, the $x$-mode changes the orbital index of the electron.
It therefore acts like the kinetic energy, but in the `orbital space', 
compare the last two terms in brackets in Eq.(\ref{five}). The only 
complication is that the rate of orbital change is not constant but
depends on the local displacement $x_{\bf r}(\tau)$. Nonetheless,
the $x$-mode and kinetic energy can be treated in the same manner,
and this is done as follows. Upon the self-multiplication of $\rho(\Delta \tau)$,
a variety of terms with different powers of $(\Delta \tau)$ appear. A term
with $n$ site changes (`kinks') and $m$ orbital changes has the weight 
\begin{eqnarray}
& (t \, \Delta \tau)^n & (\kappa\, \Delta \tau)^m 
x_{{\bf r}(\tau_1)}(\tau_1) x_{{\bf r}(\tau_2)}(\tau_2) 
\cdots x_{{\bf r}(\tau_m)}(\tau_m)  \nonumber \\
& \times \exp & \left\{ -\sum_{\bf n} \int^{\beta}_0 
\left( \frac{M \dot x^2_{\bf n}(\tau)}{2 \hbar^2} 
+ \frac{M\omega^2}{2 \hbar^2} x^2_{\bf n}(\tau) \right) d\tau \right\} .
\label{nine}
\end{eqnarray}
Here $\tau_s$ is the time of the $s$th orbital change ($s=1, \ldots, m$),
${\bf r}(\tau_s)$ is the electron position at this time, and 
$x_{{\bf r}(\tau_s)}(\tau_s)$ is the $x$-displacement at this site at 
this time. Integration over $x_{\bf n}(\tau)$ in (\ref{nine}) can now be 
performed by introducing fictitious sources, calculating the generating 
functional and differentiating it $m$ times. For any odd $m$ the result 
is zero, since the $x$ action is even under the global sign change 
$x_{\bf n}(\tau) \rightarrow -x_{\bf n}(\tau)$. (This is the above-mentioned
averaging that leads to the complete degeneracy of the two polaron bands.)
For even $m$, the time moments $\tau_s$ should be combined in pairs, 
and each pair contributes the factor $\kappa^2 G(\tau_s-\tau_{s'})$ 
with $G$ from Eq.~(\ref{eight}). Because $G$ is always positive, the
$x$-mode `favors' a large number of $G$-lines on the electron path, i.e.,
a large number of orbital changes. Such a tendency is opposite to that of the 
$y$-mode. The competition between the two modes leads to a dynamical balance
and to some average number of orbital changes per unit imaginary time.
Note also that the same function (\ref{eight}) determines both
$x$ and $y$ contributions to the density matrix, which reflects the 
equivalence of the two modes.

\begin{figure}[t]
\begin{center}
\leavevmode
\hbox{
\epsfxsize=7.4cm
\epsffile{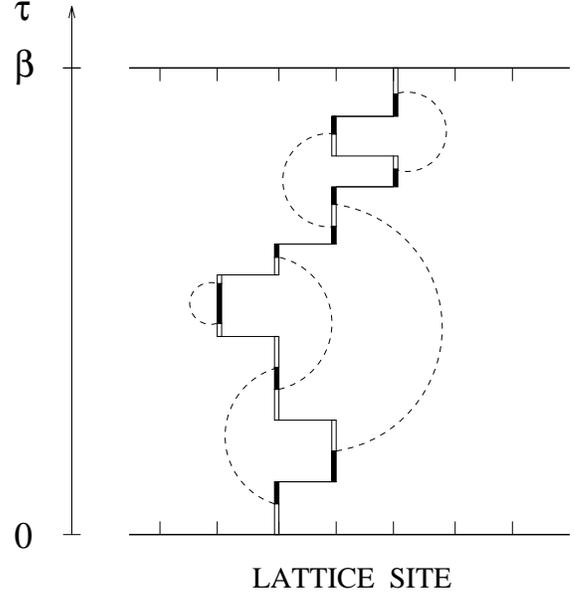}
}
\end{center}
\vspace{0.5cm}
\caption{
A typical real space - imaginary time diagram (path) with $n=8$ kinks
(5 right and 3 left ones, which makes the net shift of the path
be $\Delta {\bf r} = 5-3 = 2$ lattice sites), and $m=12$ orbital
changes (6 $G$-lines). The two different orbital states are shown
as black and white segments of the path. Each kink contributes
to path's weight the factor $t$, and each pair of orbital changes
connected by a dashed line the factor $\kappa^2 G(\tau'-\tau'')$.
Additionally, there is an overall factor $e^{A_y}$ induced by the
$y$ phonon mode. Insertion or removal of a kink at time $\tau'$
shifts the whole segment $\tau' < \tau \leq \beta$ by one lattice
site. $\Delta {\bf r}$ also changes. [Recall that $\Delta {\bf r}$
is an explicit parameter of $G$, Eq.~(\ref{eight}).] Attaching or
removing of a $G$-line at times $\tau',\tau''$ changes the orbital
index (`color') of the path at $\tau' < \tau < \tau''$. This affects
the action $A_y$, Eq.~(\ref{seven}).
}
\label{fig1}
\end{figure}

Now recall that each term (\ref{nine}) must be still integrated over all the
electron positions and orbital indices, i.e., over the time instances of 
its kinks and orbital changes. Such an integration is denoted below as
$(d\tau)^n$ and $(d\tau)^m$. In the $L \rightarrow \infty$, 
$\Delta \tau \rightarrow 0$ limit this leads to the path-integral 
representation of $\rho$ for JTP:
\begin{equation}
\rho(\beta) = \sum^{\infty}_{n=0,1, \ldots} \sum^{\infty}_{m=0,2, \ldots}
\int^{\beta}_0 \!\! \cdots \int^{\beta}_0 (d\tau)^n (d\tau)^m W_{nm} ,
\label{nineandhalf}
\end{equation}
\vspace{-0.5cm}
\begin{equation}
W_{nm} = t^n \kappa^m
\left[ \prod_{({\rm pairs \; of}\; \tau_s)} G(\tau_s-\tau_{s'}) \right] \, e^{A_y} . 
\label{ten}
\end{equation}
One cannot proceed analytically any further. However, all
the integrands are positive-definite, which suggests simulation
of $\rho$ by Monte Carlo (MC) methods. The obvious difficulty is that an 
{\em infinite series} of integrals of ever-increasing dimensionality, rather 
than just one integral, have to be evaluated. Nonetheless the recently 
developed Diagrammatic MC method \cite{Prokofiev} is capable of doing such an
integration. The method works as follows. Each $(n,m)$ term in the sum is 
represented by a diagram (path) with $n$ kinks and $m/2$ $G$-lines, 
see Fig.~\ref{fig1}. 
The MC process browses the configuration space by means of the two elementary 
subprocesses: (i) inserting and removing kinks, which changes the dimensionality
of integration by 1 ($n \rightarrow n \pm 1$); (ii) attaching and removing
$G$-lines, which changes the dimensionality by 2 ($m \rightarrow m \pm 2$).
The central idea of the method is the balance equation for the 
two subprocesses \cite{Prokofiev}. Let $N_k$ be the number of kinks of a 
given sort, $N_k \leq n$. Next, let $R(\tau')$ be the normalized
probability {\em density} with which one chooses the position for the new
kink. For example, one may decide that all the times in the $[0,\beta]$ interval
are equivalent, hence $R(\tau')= 1/\beta = {\rm const}$. In the reciprocal
removing process, one may decide that any of the $N_k+1$ existing kinks are
removed with equal probability $(N_k+1)^{-1}$. The resulting balance equation
reads:
\begin{eqnarray}
\frac{1}{\beta} \: W_{nm} P[(n,m) \rightarrow (n+1,m)] = \nonumber \\
\frac{1}{N_k+1} \: W_{n+1,m} P[(n+1,m) \rightarrow (n,m)] , 
\label{eleven}
\end{eqnarray}
from where the acceptance probabilities $P$ follow in the usual manner 
\cite{Metropolis}. The main feature to note is that the elements of the 
phase space associated with the two sides of the equation have the 
{\em same measure}: $W_{nm}$ brings $(d\tau)^{n+m}$, and $R(\tau')=1/\beta$ 
adds one more $(d\tau)$ because $R$ is a probability density. 
The measure of the right-hand-side is also $(d\tau)^{n+m+1}$. That
makes both transition probabilities be of the same order, and
renders the whole process meaningful. In the case of 
$G$-lines, let $S(\tau',\tau'')$ be the two-dimensional
probability density to attach a new $G$-line at times $\tau'$ and $\tau''$. 
While removing, each of $N_G+1$ $G$-lines may again be chosen with equal 
probability. The balance equation reads
\begin{eqnarray}
S(\tau',\tau'') \, W_{nm} P[(n,m) \rightarrow (n,m+2)] = \nonumber \\
\frac{1}{N_G+1} \: W_{n,m+2} P[(n+2,m) \rightarrow (n,m)] . 
\label{twelve}
\end{eqnarray}
Again, the measures of the phase space elements are the same because
now $S$ adds $(d\tau)^2$. A possible choice for $S$
is $S(\tau',\tau'')=(\hbar\omega/\beta) \exp{(-\hbar\omega|\tau'-\tau''|))}$. 
The MC process that follows the rules (\ref{eleven})-(\ref{twelve}) generates 
a Markov chain of paths distributed in accordance with Eq.(\ref{ten}).
On such an ensemble various polaron properties can be measured with the 
standard Metropolis rules \cite{Metropolis}. 

\begin{figure}[t]
\begin{center}
\leavevmode
\hbox{
\epsfxsize=8.4cm
\epsffile{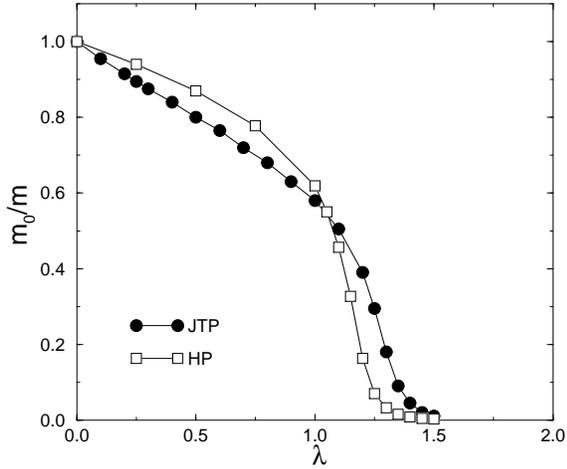}
}
\end{center}
\vspace{-0.5cm}
\caption{
Inverse effective mass of the three-dimensional JTP (circles) and HP 
(squares, adopted from [12]) for $\hbar\omega = 1.0\,t$, in units of 
$1/m_0 = 2ta^2_0/\hbar^2$, $a_0$ being the lattice constant. Statistical 
errors are smaller than the symbols.
}
\label{fig2}
\end{figure}

\begin{figure}[t]
\begin{center}
\leavevmode
\hbox{
\epsfxsize=8.4cm
\epsffile{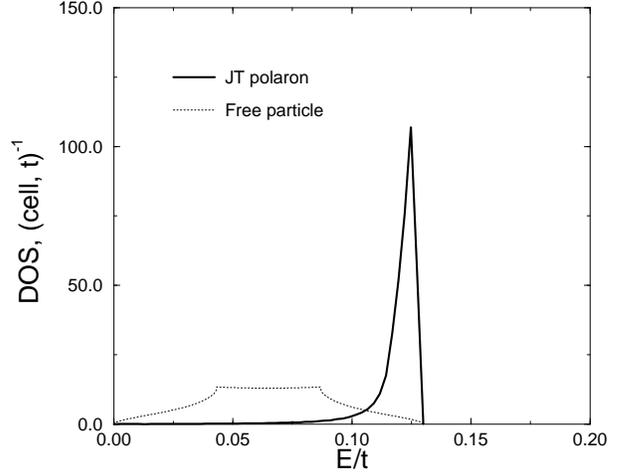}
}
\end{center}
\vspace{-0.5cm}
\caption{
Density of states of the three-dimensional JTP for $\hbar\omega = 1.0\,t$
and $\lambda=1.3$. The polaron bandwidth is $W=0.130(5)\,t$. 
The dotted line shows the free-particle DOS with the same $W$.
}
\label{fig3}
\end{figure}

One physical quantity that can be calculated with this method is the ground 
state energy of the polaron $E_0 = \langle - W^{-1}_{nm} 
(\partial W_{nm}/\partial \beta) \rangle$ \cite{DeRaedt,Kornilovitch}. 
Here we present MC data for two other properties: the effective mass and density 
of states (DOS). The mass is calculated as the diffusion coefficient of the polaron 
path, $m^{-1}_x = (\beta \hbar^2)^{-1} \langle (\Delta r)^2_x \rangle$, 
\cite{Kornilovitch}. The inverse mass of the three-dimensional JTP is shown in 
Fig.~\ref{fig2}.
After the initial weak-coupling growth $m/m_0 = 1 + {\rm const} \cdot \lambda$,
a transition to the small polaron state takes place at 
$\lambda \simeq 1.1 - 1.3$, and after that the mass increases exponentially 
with coupling. The comparison with the Holstein polaron (HP) shows that 
both masses behave similarly. The polaron spectrum is calculated with the formula 
$E_{\bf P} - E_0 = - \beta^{-1} \ln \langle \cos {\bf P} \Delta {\bf r} \rangle$,
\cite{Kornilovitch}, and DOS is obtained by integrating $E_{\bf P}$ over 
the Brillouin zone. The resulting DOS appears to be strongly distorted in 
the adiabatic regime, $\hbar\omega = 1.0\, t$, see Fig.~\ref{fig3}. 
The polaron spectrum is flat in the outer part of the Brillouin zone due 
to hybridization with dispersionless phonon modes \cite{Rashba}, 
which results in a massive peak in DOS at the top of the band. The van Hove 
singularities are invisible because they are absorbed by the peak. All these 
features are very similar to that of the Holstein polaron \cite{Kornilovitch}. 
The similarity suggests that such peculiar features of the band structure are 
common to polaron models with local interaction and dispersionless phonons in 
the adiabatic regime.

In conclusion, we have developed a path-integral representation for the Jahn-Teller 
polaron and used the Diagrammatic Monte Carlo method to simulate the series
expansion for its density matrix. The ground state energy, effective mass,
spectrum, and density of states have been calculated with no systematic errors. 
JTP has been found to be very similar to the Holstein polaron and, therefore, to 
possess some anomalous properties, in particular a strong distortion of the density
of states in the adiabatic regime, $\hbar\omega \leq t$. Such a distortion may 
be a consequence of the short-range electron phonon interaction and dispersionless 
phonon modes.

I am very grateful to Victor Kabanov for suggesting this problem and for numerous 
helpful discussions on the subject. I also benefited from conversations 
with A.\,S.\,Alexandrov, P.\,B.\,Allen, A.\,M.\,Bratkovsky, W.\,M.\,C.\,Foulkes,
and V.\,Perebeinos. This work was supported by EPSRC, grant GR/L40113, and 
by the Quantum Structures Research Initiative of Hewlett-Packard Laboratories 
(Palo Alto).

\end{document}